\documentclass[aps,prl,preprint,groupedaddress,floatfix]{revtex4}
\usepackage{graphicx}
\usepackage{epsfig}
\usepackage{amsmath}
\usepackage{amssymb}

\begin{document}

\title{Empirical Magnetic Structure Solution of\\Frustrated Spin Systems}

\author{Joseph A. M. Paddison}
\affiliation{Department of Chemistry, Inorganic Chemistry Laboratory, University of Oxford, South Parks Road, Oxford OX1 3QR, U.K.}

\author{Andrew L. Goodwin}
\email{andrew.goodwin@chem.ox.ac.uk}
\affiliation{Department of Chemistry, Inorganic Chemistry Laboratory, University of Oxford, South Parks Road, Oxford OX1 3QR, U.K.}

\date{\today}
\begin{abstract}
Frustrated magnetism plays a central role in the phenomenology of exotic quantum states. However, because the magnetic structures of frustrated 
systems are aperiodic, there has always been the problem that they 
cannot be determined using traditional crystallographic techniques.
Here we show that the magnetic component of powder neutron scattering data
is actually sufficiently information-rich to drive magnetic structure solution
for frustrated systems, including spin ices, spin liquids, and molecular magnets. Consequently, single-crystal samples are not prerequisite for detailed
characterisation of exotic magnetic states. Our methodology employs \emph{ab initio} reverse
Monte Carlo refinement, making informed use of an additional constraint
that minimises variance in local spin environments. By refining atomistic
spin configurations, we obtain at once (i) a magnetic structure
``solution'' --- \emph{i.e.} the orientation of classical spin
vectors --- (ii) the spin correlation functions, and (iii) the full
three-dimensional magnetic scattering pattern. 
\end{abstract}

\maketitle

Frustrated spin systems are characterised by the existence of a macroscopically
degenerate manifold of magnetic ground states which suppresses long-range
order \cite{Lacroix_2011}; topical examples include spin ices such
as Ho$_2$Ti$_2$O$_7$ \cite{Bramwell_2001} and 
quantum spin liquid candidates such as herbertsmithite \cite{Lee_2008,deVries_2009}.
While the absolute arrangement of spins differs amongst the degenerate
states accessible to a given system --- and hence there is no unique
or periodic ground state structure --- the states share well-defined
local correlations that distinguish their magnetic structure from
the random-spin arrangement of classical paramagnets. The importance
of understanding these correlations lies primarily in determining
the microscopic origin of exotic quantum phenomena that emerge from
or within frustrated spin systems, notably including \emph{e.g.}\ the
evolution of high-temperature superconductivity from spin liquids
\cite{Anderson_1987} and the ability of spin ices to support magnetic
monopoles \cite{Castelnovo_2008}.

Spin orientations within ordinary magnets can usually be determined
using a combination of neutron scattering experiments and crystallographic
analysis of the magnetic Bragg diffraction pattern. The suppression
of long-range spin periodicity in frustrated magnets means that no
magnetic Bragg scattering is observed. Instead the magnetic contribution
to the neutron scattering pattern is a smoothly varying function of
three-dimensional (3D) reciprocal space, whose symmetry is dictated
by the atomic lattice and whose modulation depends on the strength
and nature of the magnetic correlation functions \cite{Lacroix_2011}.
There is as yet no generic methodology for recovering the correlation
functions from the observed scattering pattern. Historically, the neutron scattering pattern anticipated from a predetermined
interaction model is calculated and compared with the experimental single crystal neutron scattering
data (see \emph{e.g.} \cite{Manuel_2009}). There are two fundamental problems with this approach. First, it relies on anticipating the nature of the interactions responsible for local
magnetic ordering. Second, it is inherently unfeasible for
the very many interesting materials for which large single crystal
samples simply do not exist.

We address both issues by developing a \emph{model-independent}
method of magnetic structure determination that exploits the information
content of the magnetic contribution $I(Q)$ to one-dimensional (1D) \emph{powder}
neutron scattering data. Drawing inspiration from the largely analogous
problem of using nuclear total scattering data to determine the
positions of atoms in disordered materials such as glasses and liquids
\cite{McGreevy_2001}, our approach is to fit powder diffraction data by refining the orientations of a configuration of semi-classical spins. We proceed
to show that the extent of information loss during spherical averaging
of the single-crystal (3D) magnetic scattering function $I(\mathbf{Q})$
is actually relatively minimal, and that the full 3D scattering pattern
is recoverable for each frustrated system of varying complexity, symmetry
and dimensionality that we explore. Moreover the corresponding real-space correlations can now be probed directly: the configurations are a ``solution'' of the frustrated magnetic structure.

For a given system, we generate a supercell of the crystallographic
unit cell and assign random spin vectors to each magnetic centre.
This spin configuration forms the starting point for a reverse Monte
Carlo (RMC) refinement, which we carry out in the usual manner \cite{Goodwin_2006,Keen_1991}. In particular, by employing starting configurations with random spin arrangements, we make no assumption regarding the form of the magnetic structure (hence  ``\emph{ab initio} structure refinement'' \cite{Juhas_2006}). 
Random spin orientation moves are generated, then accepted or rejected according
to the RMC algorithm, until the best possible fit-to-data is obtained.
We then calculate spin correlation functions from the RMC configurations
and check for evidence that all spins are in equivalent environments
(\emph{i.e.} the nearest-neighbour cross-correlation functions are
unimodal, see Fig.~S4). If such evidence is found we continue the RMC refinement
with an additional penalty term $\chi_{\textrm{Var}}^2$ that acts
to minimise variance in the six local spin correlation functions $S_{\alpha}S_{\beta}$
$(\alpha\in\{x,y\},\beta\in\{x,y,z\})$,
\begin{equation}\label{spins}
\chi_{\textrm{Var}}^2=\sum_r\sum_{\alpha,\beta}w_{\alpha\beta}\left\{ \frac{1}{N}\sum_i\left[(S_{\alpha}S_{\beta})_i(r)-\langle S_{\alpha}S_{\beta}\rangle(r)\right]^{2}\right\} ,
\end{equation}
\noindent where the $w_{\alpha\beta}$ are empirical weighting terms and the last sum is taken over all magnetic atoms $i$. In essence this
step corresponds to searching for the simplest configurations that
agree with experiment. We have shown elsewhere that the approach of
minimising local structure variance can be highly effective at improving
RMC models of disordered systems \cite{Cliffe_2010}. Finally, we
assess the quality of the models obtained by calculating the full
3D scattering function $I(\mathbf{Q})$ and also, where
appropriate, through inspection of the configurations themselves for
evidence of local spin correlations. A more complete discussion of
our particular implementation of RMC, together with the derivation
of the spin correlation functions and their variances, is given as
supporting information.

As a first case study we demonstrate the recovery of the ``ice rules'' in spin ice,
Ho$_2$Ti$_2$O$_7$ \cite{Harris_1997}, a well-studied material
in which magnetic Ho$^{3+}$ ions are arranged at the vertices of
a pyrochlore lattice of corner-sharing tetrahedra. At the simplest level, its magnetic structure involves the four moments of each Ho$_4$
tetrahedron arranged such that two point directly towards and two
directly away from its centroid: it is the macroscopic number of ways
of satisfying this local constraint for the entire structure which
leads to the observed absence of long-range order. The 1D and 3D scattering functions $I(Q)$ and $I(\mathbf{Q})$
derived from such a model are given in Fig.~1(a) and (b); these illustrate
clearly the apparent contrast in information content for which single-crystal
data are usually preferred. RMC refinement of initially-random
spin configurations yields a good
fit to data [Fig.~1(a)]. Perhaps more surprising is that the
3D scattering function $I(\mathbf{Q})$ calculated from an ensemble
of refined RMC configurations reproduces the expected pattern remarkably
well [Fig.~1(b)].

\begin{figure}
\begin{center}
\includegraphics[scale=1.6]{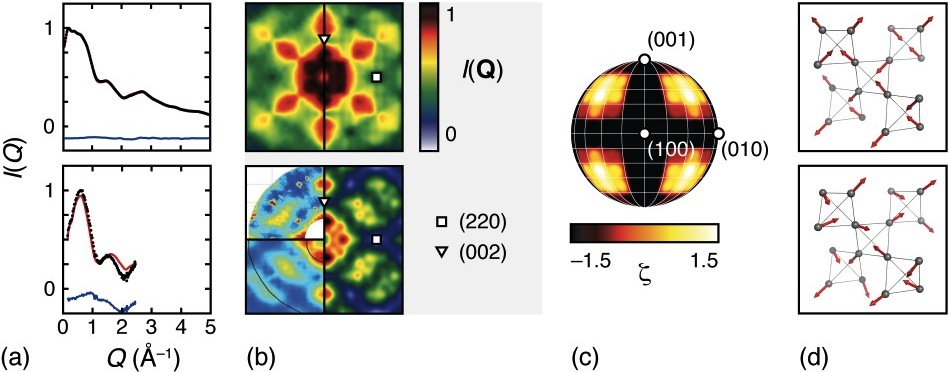}
\end{center}
\caption{\label{fig1}Spin ice, Ho$_2$Ti$_2$O$_7$. (a) Fits to neutron powder diffraction data: simulated data (top) and experimental \cite{Mirebeau_2004} data
(bottom). Input data are shown in black, RMC fit in red, and
difference (RMC$-$data) in blue. (b) Single-crystal magnetic diffuse scattering, images as follows: the ``ideal'' spin ice model described in the text (top image,
left panel); RMC-derived pattern obtained via fitting to simulated powder data (top image, right panel);
experimental \cite{Bramwell_2001b} data (bottom image, upper left-hand panel); dipolar model \cite{Bramwell_2001b}
(bottom image, lower left-hand panel); and our RMC-derived pattern obtained by fitting to neutron powder diffraction data of Ref.~\cite{Mirebeau_2004}
(bottom image, right-hand panel). (c) Real-space spin distribution
function $\zeta$ calculated from the RMC configurations (see SI), illustrating preferential
spin alignment along $\langle111\rangle$ directions. (d) Spin-ice structure (top) and a representative fragment of a RMC configuration (bottom) illustrating recovery of the spin ice rules.}
\end{figure}

Concerned that this agreement is a fortuitous result of having used
simulated data, we proceeded to use the experimental powder neutron
scattering data of Ref.~\cite{Mirebeau_2004} to drive a second
RMC refinement. The fit obtained is certainly worse [Fig.~1(a)];
indeed, the experimental data were never intended for quantitative
fitting, since they were obtained not by using polarised neutron scattering
techniques but rather as the difference between total neutron scattering
data collected above and below the spin ice ordering temperature.
Nonetheless the 3D magnetic scattering pattern calculated from our
refined RMC configurations shows a strong resemblance to the experimental
scattering data of Ref.~\cite{Bramwell_2001b} [upper left-hand panel of Fig.~1(b)]. Of particular note is that these new RMC configurations
reproduce the variation in scattering intensity attributed to dipolar
interactions [Fig.~1(b), lower left-hand panel], that is unaccounted
for in simplistic ``ice rules'' models (as above) but known to
occur experimentally \cite{Bramwell_2001b}. We note also that the
limited $Q$-range of the powder-averaged data has not affected appreciably the quality of the 3D scattering pattern
obtained from the RMC configurations. Taken together with the absence
of stringent data normalisation, this observation suggests that RMC
refinements are surprisingly robust to the inevitable limitations
of experimental data.

Access to spin configurations capable of reproducing the full $I(\mathbf{Q})$
function allows questions to be asked of the local correlations responsible
for its modulation in reciprocal space. On the simplest possible level,
it is straightforward to calculate a spin orientation distribution
function. For our experimental-data-driven RMC refinement of Ho$_2$Ti$_2$O$_7$
such a calculation clearly reflects the preferential alignment along
the $\langle111\rangle$ directions expected for spin ice [Fig.~1(c)].
Furthermore, inspection of any of the refined RMC configurations (one
of which is included as supporting information) betrays the ice rules
themselves: we find that approximately 95\% of Ho$_4$ tetrahedra
obey the ``2-in-2-out'' rule [Fig.~1(d)]. To the best of
our knowledge, this represents the first model-independent, data-driven
``solution'' of the spin ice structure.

We proceeded to test whether our RMC approach enjoyed similar success
across a range of frustrated magnets of varying topology, form of magnetic
interaction, lattice symmetry and dimensionality, focussing our
choices on systems of long-standing interest and/or particular currency within the condensed matter
physics community. We have assembled a visual summary of our results
--- for Heisenberg pyrochlores \cite{Moessner_1998} (including the
``hexagonal spin cluster'' spinel ZnCr$_2$O$_4$ \cite{Lee_2002,Conlon_2010}),
gadolinium gallium garnet (GGG) \cite{Petrenko_1998}, the hyperkagome
Na$_4$Ir$_3$O$_8$ \cite{Okamoto_2007,Hopkinson_2007}, the
hexagonal extended kagome YBaCo$_4$O$_7$ \cite{Manuel_2009},
and \emph{XY }and Ising 2D kagome lattices \cite{Wills_2002,Chalker_1992}
--- in Fig.~2, with full details of these refinements and a further
parallel refinement of GGG driven by experimental data \cite{Mirebeau_2004}
given as supporting information [Fig.~S6]. In each case we find that the 3D
magnetic scattering function is almost entirely recoverable from the
1D powder-averaged data. Moreover, despite showing only specific high-symmetry
cuts of reciprocal space in Fig.~2, one has access to the full volume
of reciprocal space from the RMC configurations. Historically this has always been difficult,
even with access to large single crystal samples, by virtue of the
time-consuming experimental process of assembling 3D reciprocal space maps from 2D data sets.

\begin{figure}
\begin{center}
\includegraphics[scale=1.6]{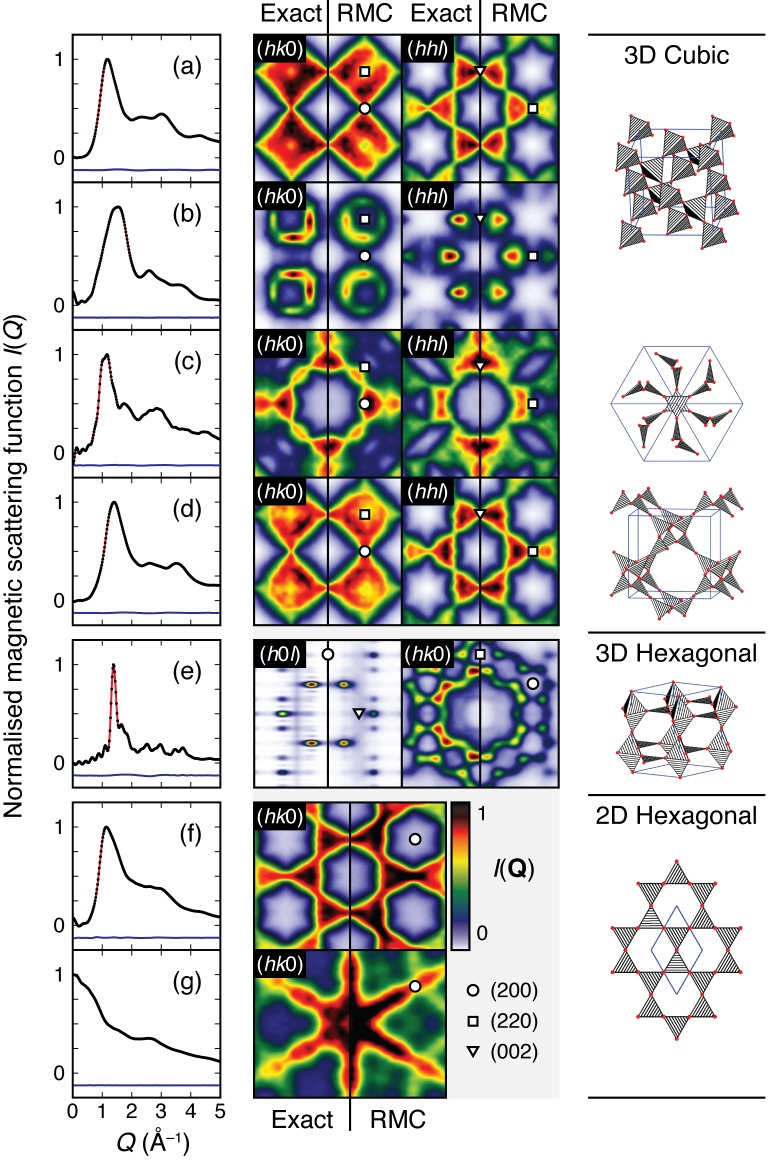}
\end{center}
\caption{\label{fig2}Summary of results for frustrated magnets. (a) Heisenberg pyrochlore
antiferromagnet; (b) ``hexagonal spin cluster'' spinel ZnCr$_2$O$_4$;
(c) gadolinium gallium garnet (GGG); (d) hyperkagome Na$_4$Ir$_3$O$_8$;
(e) extended kagome YBaCo$_4$O$_7$; (f) kagome \emph{XY} system;
(g) kagome Ising system. The left-hand column shows the RMC fit to
simulated powder diffraction data (calculation details given as SI).
The central column shows 2D cuts of the magnetic scattering function I$(\mathbf Q)$: those
labelled ``Exact'' are calculated from the model spin
configurations used to generate the simulated powder diffraction data; those labelled ``RMC''  are calculated from the RMC spin configurations
obtained by fitting to these powder data. Diagrams of the corresponding
frustrated lattices are shown in the right-hand column.}
\end{figure}

Local magnetic order is less well defined in each of these particular frustrated magnets than is the case for spin ice; consequently, direct inspection
of the configurations is not always so informative. We do, however,
recover the signature of spin planarity in the 2D systems despite employing
3D Heisenberg spins in the RMC refinement [Fig.~S5]. For all seven
systems it is possible to extract the spin correlation functions,
which reflect the (anti)ferromagnetic nature and correlation length
of spin interactions [Fig.~S3]; we note that in principle these could
perhaps be used further to drive independent estimation of spin interaction
parameters \cite{Almarza_2003}.

Extending our methodology to a third distinct class of frustrated spin
system, we consider as a final case study the single-molecule magnet
Mo$_{72}$Fe$_{30}$ \cite{Muller_1999}. The magnetic Fe$^{3+}$ ions
occupy the corners of an icosidodecahedron, forming a network of corner-sharing
triangles that is a zero-dimensional analogue of the kagome lattice.
Its frustrated magnetic structure has been described in terms of a
``3-sublattice'' model \cite{Muller_2001,Axenovich_2001}, in
which three groups of ten collinear spins point along axes separated
by angles of 120$^\circ$ [Fig.~3(a)]; the absolute orientation of the vectors is
not uniquely determined. In this way the vertices of each triangle in the icosidodecahedron are decorated by one spin from each
sublattice.  Once again our RMC refinements, driven by simulated neutron powder diffraction data, consistently produced spin configurations that shared
the general features of this magnetic order, with only the absolute
orientation of the three ordering vectors differing between configurations
[Fig.~3(b)].

\begin{figure}
\begin{center}
\includegraphics[scale=1.6]{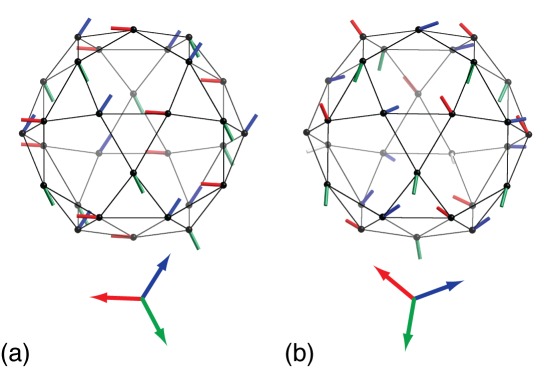}
\end{center}
\caption{\label{fig3}Mo$_{72}$Fe$_{30}$ single-molecule magnet. (a) The previously-described spin structure \cite{Muller_2001,Axenovich_2001};
(b) a typical RMC magnetic structure solution obtained from simulated powder neutron scattering data. Red, green and blue vectors indicate spin orientations of
the three sublattices.}
\end{figure}

The use of RMC methods in the solution of \emph{non-magnetic} disordered structures has a chequered history \cite{Biswas_2004}; in this light the apparent success for magnetic structure solution is perhaps surprising. For frustrated magnets, however, the problem of structure solution will be heavily constrained by knowledge of the positions of magnetic atoms. We also find that the inclusion of the local spin invariance term of Eq.~\eqref{spins} plays a crucial role for systems where the degree of local order is strong; for example, repeating our spin ice refinements without this term we obtain configurations that are convincing neither in their reproduction of the $I(\mathbf Q)$ scattering function [Fig.~S2] nor in their adherence to the spin ice rules [Table~S3]. Our results suggest two particularly significant opportunities for further study. First, direct investigation of magnetic correlations in systems not available as single crystals --- such as the quantum spin liquid $\kappa$-(BEDT-TTF)$_2$Cu$_2$(CN)$_3$ \cite{Pratt_2011} --- becomes a viable prospect. Second, one can now develop microscopic pictures of spin structures where there exists a signature of strongly correlated magnetism, \emph{e.g.}\  ``spin stripes'' in high-temperature superconductors \cite{Boothroyd_2011}.  Consequently, the methods developed here offer a valuable new tool with which to explore the magnetic structures of frustrated systems associated with quantum phenomena of fundamental importance.

We gratefully acknowledge valuable discussions with D. A. Drabold, D. A. Keen, M. J. Cliffe, L. C. Chapon, J. R. Stewart and S. T. Bramwell, and funding from the EPSRC (UK) under grant EP/G004528/2.

\end{document}